# Octapod-shaped CdSe nanocrystals hosting Pt with high-mass activity for the hydrogen evolution reaction

Leyla Najafi,[‡a] Sebastiano Bellani,[‡a] Andrea Castelli,[b] Milena P. Arciniegas,[b] Rosaria Brescia,[c] Reinier Oropesa-Nuñez,[d] Beatriz Martín-García,[a] Michele Serri,[a] Filippo Drago,[b] Liberato Manna[b] and Francesco Bonaccorso[*a,d]

[a] *Graphene Labs, Istituto Italiano di Tecnologia, via Morego 30, 16163, Genova, Italy.*

[b] *Nanochemistry Department, Istituto Italiano di Tecnologia, via Morego 30, 16163 Genova, Italy*

[c] *Electron Microscopy Facility, Istituto Italiano di Tecnologia, via Morego 30, 16163 Genova, Italy*

[d] *BeDimensional Spa., Via Albisola 121, 16163 Genova, Italy*

**ABSTRACT:** The design of efficient electrocatalysts for electrochemical water splitting with minimal amount of precious metal is crucial to attain renewable and sustainable energy conversion. Here, we report the use of a network of CdSe branched colloidal nanocrystals, made of a CdSe core and eight CdSe pods (so-called octapods), able to host on their pods Pt particles, and thus catalyzing water splitting reactions. Thanks to the octapod shape, the resulting Pt-hosting network is mechanically trapped onto carbon nanotube buckypaper, providing mechanically flexible and binder-free electrodes. We found that such hierarchical configuration maximizes the mass activity and the utilization efficiency of Pt for the hydrogen evolution reaction (HER). At a potential of -0.15 V *vs.* reversible hydrogen electrode, the Pt/octapod network-based electrodes display a Pt mass activity on the HER of ~166 A mg$^{-1}$ and ~42 A mg$^{-1}$ in acidic and alkaline media, respectively. These values correspond to turnover frequencies of ~168 s$^{-1}$ and ~42 s$^{-1}$, respectively, which are in that order 14 and 21 times higher compared to commercially available Pt/C benchmarks. The strong chemical and mechanical interactions between the Pt and the octapod surface, along with pod-aided adhesion of the Pt/octapod network to the buckypaper, result in a long-term durability (>20 h) of the HER-activity in both media. These results experimentally prove that the exploitation of our network of branched nanocrystals hosting Pt particles can circumvent the durability issues of the catalysts while adopting either ultralow Pt loadings or benchmarking carbon-supported Pt nanocrystals. Our work opens up prospects for using porous networks made by branched nanocrystals as catalysts with ultralow amount of noble metals and controlled catalytic properties.

## INTRODUCTION

Molecular hydrogen (H$_2$) production *via* electrochemical water splitting is considered the cleanest route toward the realization of the so-named "hydrogen economy",[1] which envisions the use of H$_2$ as unique renewable and sustainable energy source.[2] To tackle this challenge, the Pt-group elements have been the most effective choice to efficiently catalyse the hydrogen evolution reaction (HER)[3,4], *i.e.*:

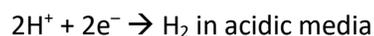

2H$^+$ + 2e$^-$ → H$_2$ in acidic media

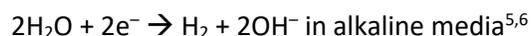

2H$_2$O + 2e$^-$ → H$_2$ + 2OH$^-$ in alkaline media[5,6]

However, the high cost and the scarcity of these materials (*e.g.*, for Pt, > US$ 40 g$^{-1}$ and < 0.005 ppm, respectively)[7,8] hamper the market entry of the electrochemical water splitting technologies,[9,10] which are still far from conveniently replacing the diminishing fossil fuels.[11,12] To overcome these hurdles, catalysts based on non-precious materials have been extensively reported.[13–15] However, they typically suffer from remarkable overpotentials (in the order of hundreds mV) or poor stability (hour-/day-time scale) in acidic media,[16,17] impeding their application in inexpensive efficient acidic solid polymer electrolytes.[18–20] Therefore, Pt remains as a universal benchmark for carrying out HER. Major efforts have been pursued to maximize both its surface catalytic activity and surface-to-bulk atomic ratio in order to boost its mass activity ($MA_{Pt}$). The $MA_{Pt}$ is defined as the Pt mass-normalized current, which is attributed to a specific reaction at a given potential.[16,21–33] To expose nearly all of the Pt atoms to the electrolyte, at which HER can be performed, Pt atoms,

nanocrystals[21,22,23–30,34] or monolayers[31,32] are typically embedded/deposited into/onto appropriate supports, as it is done in benchmark catalysts, such as Vulcan XC-72-supported Pt nanocrystals (Pt/C). These supports are preferably high-surface area and/or electrically conductive scaffolds.[33,35,36] However, Pt nanocrystals/monolayers often suffer from instability caused by Pt dissolution or delamination during the water splitting reactions.[37,38] The preparation of single Pt atom catalysts, *e.g.*, Pt atoms on $CeO_2$,[39] $Al_2O_3$,[40] TiN,[41] TiC,[41,42] carbon,[43] MXenes,[22] is challenging since the single Pt atoms tend to coalesce into clusters during the catalyst synthesis and especially during the electrocatalytic processes,[44,45] quickly degrading the theoretical maximum $MA_{Pt}$.[46] Therefore, it is essential to isolate and immobilize Pt atoms and to strengthen the adhesion and chemical stability of the Pt nanocrystals/monolayers, to prevent the aforementioned issues. This is achievable through the Pt localization into/onto the crystal lattice of an hosting matrix/substrate through controlled chemical interactions.[22,47–50] The structural mismatch at the Pt/support interface can be exploited to control the lattice strain of Pt nanocrystals/monolayers, tuning their catalytic properties.[51–53] The support can also influence the formation of intrinsically strained multiple twinned structures, which show superior catalytic activity compared to the one of single crystals.[54,55] Furthermore, the manipulation of the coordination environments of Pt enables tuning of their electronic structures to accelerate the kinetics of targeted electrochemical reactions.[24,56,57]

In this work, we: 1) exploit the peculiar shape of a class of branched nanocrystals formed by eight CdSe pods protruding from a CdSe core, namely CdSe octapods (CdSe-OCPs);[58] 2) decorate their surface with Pt (Pt@CdSe OCPs); 3) assemble the hierarchical nanostructures in the form of a porous network onto carbon nanotube (CNT) buckypaper as the current collector (CNT/Pt@CdSe-OCPs), and 4) investigate the resulting system as electrode for water splitting reactions. The CdSe OCP are synthesized by a colloidal seeded-growth approach.[59,60] The Pt@CdSe-OCPs are prepared by nucleating multiply twinned and strained Pt nanocrystals onto the sidewalls of the CdSe-OCP pods. Furthermore, they intrinsically maximize the distance between Pt nanocrystals decorating the different arms of each octapod, creating an isolation-induced stabilization effect of Pt, *i.e.*, preventing the Pt to coalesce. Consequently, ultralow amounts of Pt (in the order of few µg $cm^{-2}$) are stably hosted in the Pt@CdSe-OCP network, which is simply obtained through vacuum filtration of the Pt@CdSe-OCP dispersion onto CNT buckypaper. The Pt@CdSe-OCPs, thanks to their branched shape, self-assemble into a porous network when deposited on a substrate.[61–65] Specifically for our case, the Pt@CdSe-OCP network can strongly adhere to CNT buckypaper without using any polymeric binding agent (*e.g.*, tetrafluoroethylene-perfluoro-3,6-dioxa-4-methyl-7-octenesulfonic acid copolymer –Nafion– typically used for HER-catalyst in acidic media, including Pt/C),[66,67] since the pods effectively grab the CNT tangle. The hierarchical electrodes maximize the catalytic properties and the utilization efficiency of the Pt, which show a superior $MA_{Pt}$ compared to Pt/C reference, by more than one order of magnitude, while circumventing durability issues of the catalysts adopting ultralow Pt loadings.

**EXPERIMENTAL SECTION**

The chemicals, the syntheses of $Cu_{2-x}Se$ seeds, CdSe-OCPs and CdSe quantum dots (QDs), the debundling of CNTs,[68,69,70] and material/electrode characterization methods are reported in the Supporting Information.

**Synthesis of Pt@CdSe-OCPs**.[71] In a three-neck flask 21.5 mg of 1,2-hexadecanediol, 0.1 mL of oleic acid and 0.1 mL of oleylamine (OLAM) were dissolved in 5 mL of diphenyl ether and degassed for 1h at 80 °C under stirring. The solution was then heated up to 225 °C under $N_2$ flow before injecting an anhydrous mixture of 0.5 mL dichlorobenzene, 1.8 mL of octapods in toluene (for a total of 12.5 mg of CdSe) and 0.5 mg of Pt(II) acetylacetonate (Pt(acac)$_2$). The reaction was run for 10 min before cooling it down to room temperature. When the temperature dropped below 100 °C, 2.5 mL of anhydrous toluene were added to the resulting product. The resulting dispersion was then washed twice using a mixture of ethanol and methanol and centrifuged to remove the excess of free-ligands.

**Preparation of the electrodes.** The electrodes were prepared by sequentially depositing CNTs and CdSe-OCPs, or Pt@Cd-Se-OCPs, or CdSe-QDs from their dispersions onto nylon membranes (Whatman® membrane filters nylon, 0.2 µm pore size) by means of a vacuum filtration process.[72] The material mass loading was 1.31 mg cm$^{-2}$ for CNTs and 38 µg cm$^{-2}$ for the other catalytic materials. The electrode area was 3.8 cm$^2$. Subsequently, the electrodes were washed with 10 mL of isopropyl alcohol (IPA) to remove residual ligands from Pt@CdSe-OCP surface, as well as residuals of N-methyl-2-pyrrolidone (NMP), which was used to disperse CNTs. Comparative surface analysis performed on the as-prepared and purified Pt@CdSe-OCP *via* Fourier-transformed infrared spectroscopy –FTIR– (**Figure S1**) confirms the removal of the native phosphorous-based ligands from the Pt@CdSe-OCP surfaces by a strong reduction on the signal of the phosphorous-based ligands.[64] Lastly, the electrodes were dried overnight at room temperature. To prepare Pt/C electrodes, a dispersion of Pt/C was produced by dissolving 4 mg of Pt/C and 80 µL of Nafion solution in 1 mL of 1:4 v/v ethanol/water. Subsequently, the Pt/C dispersion was drop-cast onto glassy carbon (GC) substrates (Sigma Aldrich). The Pt/C mass loading of the electrodes was 0.262 mg cm$^{-2}$, in agreement with previous reported protocols.[73,74]

## RESULTS AND DISCUSSION

**Synthesis and characterization of the Pt@CdSe-OCPs.** The synthesis of Pt@CdSe-OCPs is briefly illustrated in the Supporting Information (**Scheme S1**). The anisotropic octapod shape is obtained through a seeded-growth approach.[59,60] Specifically, $Cu_{2-x}Se$ nanocrystals, with truncated octahedra shape and having cubic berzelianite crystal structure are firstly synthesized. These, then, undergo Cd-cation exchange (leading to CdSe cores with cubic sphalerite structure) and pod growth (wurtzite-type shell) on a second step, delivering CdSe-OCPs.[64,75,76] Finally, the CdSe-OCPs serve as substrate for the heterogeneous nucleation of Pt domains in the third step (see Experimental Section for details).[71] The amount of Pt grown on chalcogenide branched nanoparticles can be easily tuned by varying the initial concentration of metal precursor,[71,77] beyond the choice of surface ligands.[71] However, the increase of the Pt amount might not be beneficial for the overall electrocatalytic performance of the system, since Pt nanocrystals can increase their size, reducing their surface-to-bulk atomic ratio.[71] In this regard, the ratio between Pt precursors and CdSe-OCPs used in this work (4%) was similar to the one previously used to obtain nm-size Pt nanocrystals.[71]

The use of $Cu_{2-x}Se$ nanocrystals allows us to obtain, as seed, a faceted particle > 10 nm with cubic crystal structure, which can readily undergo cation exchange with Cd, without transitioning to a hexagonal (wurtzite-type) lattice.[64,78,79] **Figure 1a** illustrates a sketch of an octapod, indicating the relevant octapod dimension: the tip-to-tip arm length (*L*), the arm diameter (*D*), and the actual arm length (*L$_P$*). *L$_P$* length is defined as $L_P = L/2 - R_{seed}$, where $R_{seed}$ indicate the radius of the CdSe core. $R_{seed}$ is approx. 12.2 ± 1.5 nm as evaluated by detailed analysis of bright-field transmission electron microscopy (BFTEM) images of Cd-cation exchanged $Cu_{2-x}Se$ seeds (**Figure 1b**), *i.e.*, CdSe-QDs (see Experimental Section for details). **Figure 1c** shows a BFTEM image of two representative CdSe-OCPs before being decorated with Pt nanocrystals. The CdSe-OCP dimensions are: *L* = 57 ± 7 nm; *D* = 12.8 ± 1.9 nm; and *L$_P$* = 22 ± 4 nm (calculated from the statistical analysis of BFTEM images showing CdSe-OCP ensembles, **Figure S2**), resulting in an aspect ratio *L/D* = 4.6 ± 0.9. The structural properties of the CdSe-OCPs were evaluated by X-ray diffraction (XRD) analysis. As mentioned, CdSe can appear in two crystalline forms, *i.e.,* the cubic (sphalerite or zinc-blende, space group F-43m) and the hexagonal (wurtzite, space group P63 mc) structures.[80,81] **Figure 1d** shows the XRD pattern of CdSe-OCPs, in comparison to that of CdSe-QDs, as representative of the crystalline structure of the CdSe-OCP core. These latter evidence a pure sphalerite phase (ICDD card no. 98-018-6011),[75] while the CdSe-OCPs clearly shows also the diffraction peaks related to the wurtzite phase (ICDD card no. 98-041-5786), which is attributed to the octapod arms grown on the {111} facets of the CdSe cores.[82] **Figure 1e** shows a high-angle annular dark-field scanning TEM (HAADF-STEM) image of few Pt@CdSe-OCPs, exhibiting a morphology that is similar to the one of the native CdSe-OCPs, except for the presence of Pt nanocrystals onto the octapod arms. More in detail, the fast Fourier transform (FFT) analysis of high resolution TEM (HRTEM) images (**Figure S3**) indicates that the pods are wurtzite-type (hexagonal) CdSe (ICSD 415785) (similarly to the pristine CdSe-OCP), with <001> as elongation direction. The HRTEM analysis of a Pt nanocrystal (**Figure 1f**) evidences that the latter has a twinned structure. The energy dispersive X-ray spectroscopy in STEM (STEM-EDS) analysis shows a low content of Pt (Pt/Cd atomic ratio lower than 5%), which translates into a noisy map, except in regions where the nm-size Pt nanocrystals are positioned (**Figure 1g**). The chemical composition of the Pt@CdSe-OCP surface, as well as the binding status of the composing elements, were investigated through X-ray photoelectron spectroscopy (XPS) measurements. The XPS data obtained for the native CdSe-OCPs are reported in the Supporting Information (**Figure S4**), together with their discussion. For the Pt@CdSe-OCPs, the Cd 3*d* spectrum (**Figure 1h**) features two peaks positioned at 405.1 eV and ~411.8 eV, which are assigned to Cd $3d_{5/2}$ and Cd $3d_{3/2}$ bands, respectively, of the $Cd^{2+}$ state of the CdSe.[83] In the Se 3*d* spectrum (**Figure 1i**), the bands peaking at 53.9 eV and 54.8 eV are attributed to the Se $3d_{5/2}$ and Se $3d_{3/2}$, respectively, of the Se moiety in the CdSe. A negligible peak at 59.0 eV is ascribed to a marginal atomic content (0.8% of total Se) of $Se^{4+}$ state in $SeO_2$.[84,85] The Pt 4*f* signal is clearly visible in Pt@CdSe-OCPs, although it overlaps with a broad energy loss structure observed in the CdSe-OCPs (**Figure S5**). **Figure 1j** reports the Pt 4*f* spectrum after subtraction of the CdSe-OCP background (**Figure S4c**), with peaks at 71.5 eV and 74.8 eV attributed to Pt $4f_{7/2}$ and Pt $4f_{5/2}$ core levels, respectively. These peak binding energies are slightly shifted if compared to those observed for metallic $Pt^0$ state that is expected from Pt nanocrystals (*i.e.*, 71.2 eV and 74.5 eV).[16,23,57] Moreover, the Pt 4*f* peaks have asymmetric shapes, which can be ascribed to the presence of Pt atoms with partial positive charges ($Pt^{+\delta}$).[57,86–88] The latter suggest a strong chemical bonding between Pt and the underlying material,[30,57,86–91] *i.e.*, CdSe-OCPs. The metallic behaviour of Pt nanocrystals could also be the cause of the asymmetric shapes of Pt 4*f* peaks (see **Figure S6**).[92] However, the optimal fit shown in **Figure 1j** is obtained using symmetric peaks and introducing a fraction (~11%) of $Pt^{+\delta}$ contributing with a doublet peaking at higher binding energy (73.5 eV and 76.8 eV) compared to those of $Pt^0$ doublet. Overall, the energy shift of the $Pt^0$ doublet peaks and the possible presence $Pt^{+\delta}$ species suggest a chemical interaction between Pt and the CdSe-OCP surface.[57,86–88] The elemental analysis estimates a Cd:Se atomic ratio of ~1.3, indicating the presence of a Cd-rich surface. This might alter the electron density in the CdSe lattice, allowing the atomic Pt to stably fit in defective sites, whereas the CdSe surrounding structure can also minimize the Pt dissolution (see below the electrochemical characterization).[93] The atomic content of Pt onto the Pt@CdSe-OCPs surface, determined by XPS, is 2.2% of total Cd, Se and Pt (excluding C and O attributed to (free-)ligands). However, it is important to highlight that, during the Pt@CdSe-OCP synthesis, Pt is specifically grown onto the CdSe-OCPs, while the inner structure of the CdSe-OCPs is reasonably unaltered (as indicated by the XRD and the HAADF-STEM analyses). Therefore, the actual at% of Pt in the entire Pt@CdSe-OCPs (accounting of its inner structure) is significantly lower than the value estimated by the surface sensitive XPS technique. Inductively coupled plasma optical emission spectroscopy (ICP-EOS) and ICP-mass spectroscopy (ICP-MS) measurements of Pt@CdSe-OCP dispersion were carried out to estimate the mass/atomic content of the Cd, Se and Pt on the entire Pt@CdSe-OCPs. Interestingly, the data revealed an atomic content of Pt relatively to Cd, Se and Pt as low as 0.71% (corresponding to a mass content relatively to the mass of Cd, Se and Pt of 0.76%). An excess of Cd relatively to Se was also detected (Cd:Se atomic ratio of ~1.2), slightly reduced compared to the one detected by XPS on the Pt@CdSe-OCP surface.

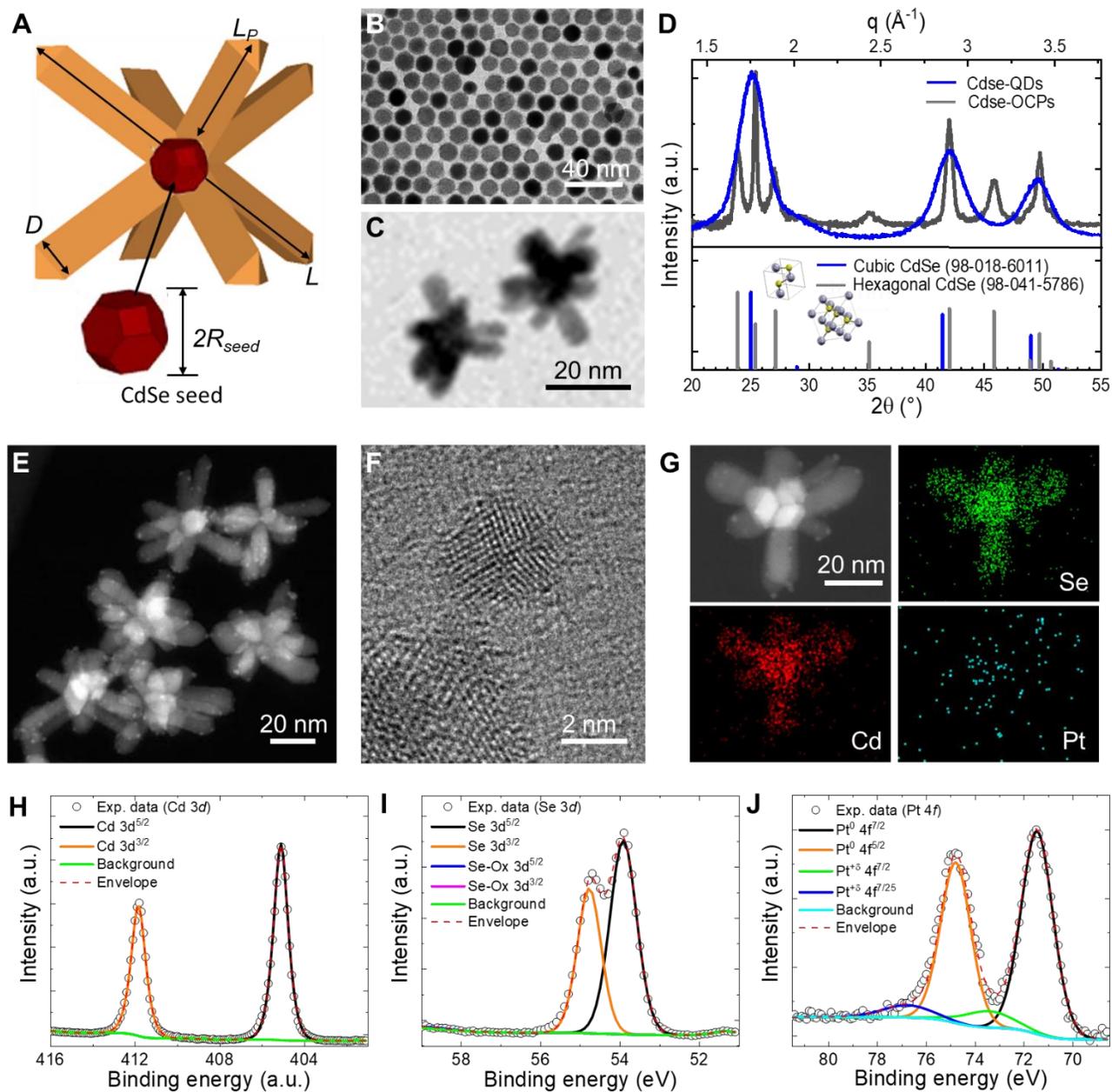

**Figure 1.** Morphological, structural and chemical characterization of the as-synthetized CdSe-OCPs and Pt@CdSe-OCPs. (A) Sketch illustrating the geometrical parameters of a CdSe-OCPs. $L$ and $D$ represent the tip-to-tip arm length and the arm diameter, respectively. The actual arm length is represented by $L_P$. The truncated octahedral CdSe core is indicated in dark red and has a radius of $R_{seed}$. (B) BFTEM image of representative Cd-cation exchanged $Cu_{2-x}Se$ nanocrystals, *i.e.*, CdSe-QDs, obtained from CdSe-OCPs synthesis without additional Se precursor. (C) BFTEM image of two representative CdSe-OCPs. (D) XRD patterns of CdSe-QDs and CdSe-OCPs (top), together with the reference XRD patterns (bottom). (E) HAADF-STEM of representative Pt@CdSe-OCPs. Pt particles are visible as brighter particles on the octapods arms. (F) HRTEM image of a multiply twinned and strained Pt particle, as grown onto CdSe-OCPs during the Pt@CdSe-OCP synthesis. (G) HAADF-STEM image of an isolated Pt@CdSe-OCPs and the corresponding STEM-EDS maps for Se (green), Cd (red) and Pt (cyan). (H) Cd 3$d$, (I) Se 3$d$ and Pt 4$f$ XPS spectra of Pt@CdSe-OCPs, together with their deconvolutions.

**Electrochemical characterization the of Pt@CdSe-OCP-based electrodes.** The HER-activity of the Pt@CdSe-OCPs was evaluated in a three-electrode configuration system by linear sweep voltammetry (LSV) measurements, with a scan rate of 10 mV s$^{-1}$ and at room temperature, in both acidic (0.5 M $H_2SO_4$) ad alkaline media (1 M KOH). The electrodes were prepared through a room temperature vacuum filtration of the produced catalyst dispersions (catalyst mass loading = 38 µg cm$^{-2}$) onto a CNT buckypaper (CNT mass loading = 1.31 mg cm$^{-2}$). Thanks to their branched structure, the Pt@CdSe-OCPs self-assemble in a network during their deposition on the selected substrate. In addition, the as-formed network strongly adheres to the CNT buckypaper (electrode named CNT/Pt@CdSe-OCPs), without resorting to any binding agent (*e.g.*, tetrafluoroethylene-perfluoro-3,6-dioxa-4-methyl-7-octenesulfonic acid copolymer –Nafion–, as typically adopted for traditional catalysts,[66] including Pt/C[67]), since the pods effectively grab the CNT tangle (**Figure 2a**). On the contrary, the CdSe-QDs can pass through the CNT tangle, leading to catalytic material losses, as well as a weak adhesion of the dot-shaped catalysts to the buckypaper (electrode named CNT/CdSe-QDs). The detailed description of the preparation of the electrodes is reported in the Experimental Section. **Figure 2b** shows a photograph of a representative CNT/Pt@CdSe-OCPs, which was bent to illustrate its mechanical flexibility. **Figure 2c,d** show representative top-view SEM images of the CNT/Pt@CdSe-OCPs, revealing a homogeneous Pt@CdSe-OCP network adhered to the wavy CNT buckypaper, which form a mesoporous tangle-like layer (**Figure S7**). Differently, the top-view SEM image of the CNT/CdSe-QDs (**Figure 2e**) shows that the CdSe-QDs are largely dispersed within the CNT tangle, which is not effective for trapping dot-shaped nanocrystals, as sketched in **Figure 2a**. The elemental analysis of CNT/Pt@CdSe-OCPs was performed through SEM-EDS measurements (**Figure S8**) revealing a Cd:Se atomic ratio of ~1.1. This indicates the presence of a Cd-rich surface, in agreement with XPS and ICP-OES/MS analyses. Furthermore, no Pt has been detected, confirming the ultralow Pt mass loading deduced by ICP-MS measurements and the Pt@CdSe-OCP mass loading (38 μg cm$^{-2}$).

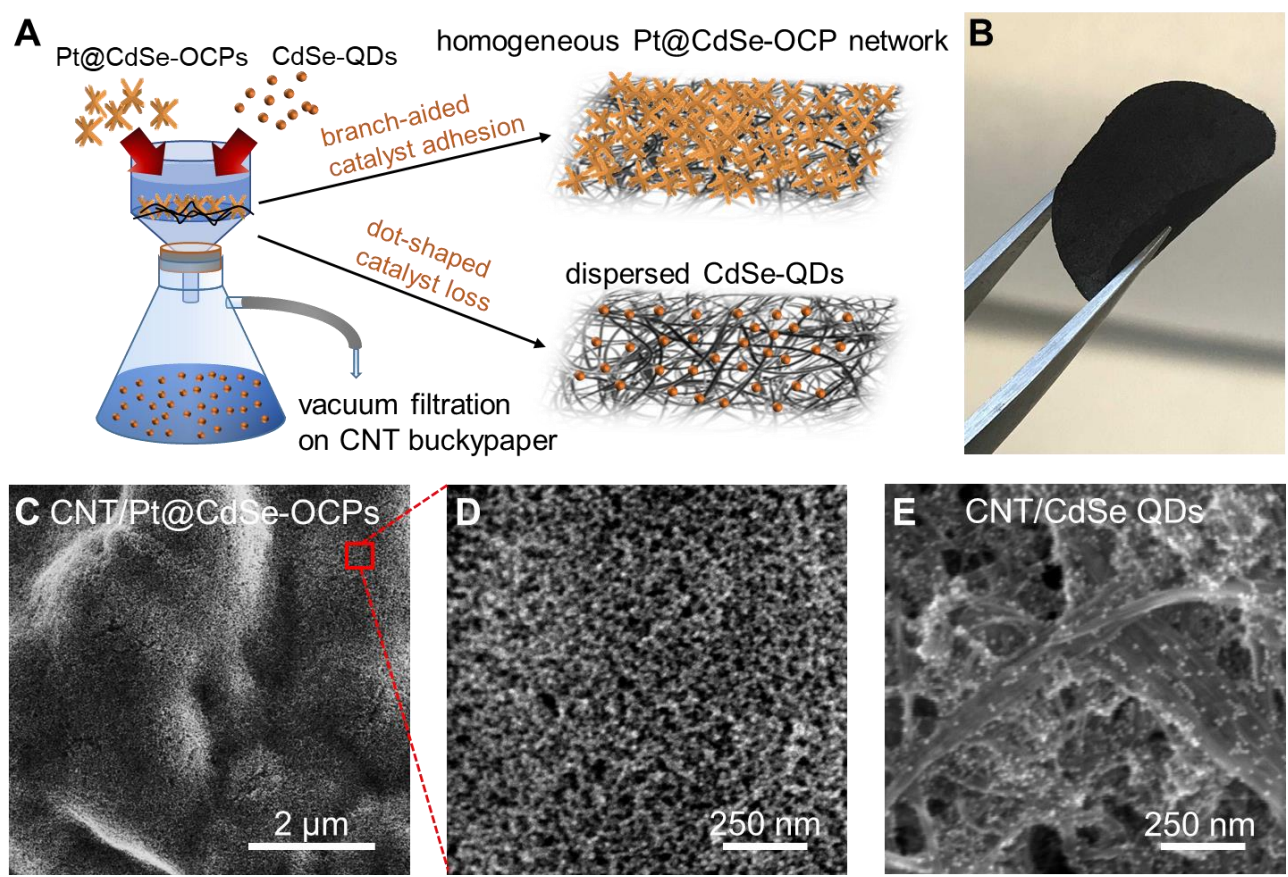

**Figure 2.** (A) Illustration of the electrode fabrication process by room temperature vacuum filtration of the catalyst dispersions onto CNT buckypaper. A homogeneous Pt@CdSe-OCP film is deposited over CNTs (electrode named CNT/Pt@CdSe-OCPs), since the branched catalyst structure is naturally trapped by the CNTs. On the other hand, CdSe-QDs can pass through the CNTs (resulting electrode named CNT/CdSe-QDs), leading to catalytic material losses, as well as a weak adhesion of the dot-shaped catalysts to the buckypaper. (B) A photograph of a flexible self-standing electrode produced by room-temperature sequential vacuum filtration of SWCNTs and Pt@CdSe-OCP dispersions onto nylon membrane. (C) Top-view SEM image of the CNT/Pt@CdSe-OCPs and (D) its high-magnification SEM image of the red-marked area. (E) Top-view SEM image of the CNT/CdSe-QDs, showing the bundle-like morphology of the CNT buckypaper

In order to provide a reliable comparison between the catalytic performance of our electrodes with those reported in literature, we specifically focused on the analysis of $MA_{Pt}$ (instead of electrochemically active surface area ($ECSA$)-activity) to evaluate the intrinsic HER-activity of Pt. In fact, since highly porous substrates (*i.e.*, CNT buckypapers) were used to support our catalytic films, the evaluation of the $ECSA$ of our catalytic films through traditional methods (*e.g.*, double-layer capacitance –$C_{dl}$– measurements through analysis of cyclic voltammetry –CV– scans at different potential scan rates in a non-faradaic region)[94] can lead to overestimated values, which means an underestimation of the catalytic performance.[95,96] As shown in **Figure S9**, the CNT buckypapers display a $C_{dl}$ as high as ~790 mF cm$^{-2}$, which inevitably impedes an accurate determination of the $C_{dl}$ of our catalytic films deposited onto CNT buckypapers. **Figure 3a** shows the cathodic LSV scan of the CNT/Pt@CdSe-OCPs, together with the one measured for a commercial Pt/C benchmark (produced accordingly to previously reported protocols[74,73], see Experimental Section). The current density in the LVS plots was normalized to the mass loading of Pt (0.29 µg cm$^{-2}$ for CNT/Pt@CdSe-OCPs; 26.2 µg cm$^{-2}$ for Pt/C) (see Supporting Information, Equation (S3)), thus reporting the –$MA_{Pt}$ *vs.* potential. It is worth pointing out that, as shown in **Figure S10**, the bare CNT/CdSe-OCPs do not exhibit relevant HER-activity in the potential windows examined in our work and reported in **Figure 3a**. This means that the estimated $MA_{Pt}$ for the CNT/Pt@CdSe-OCPs can be rightly attributed to the Pt interacting with the CdSe-OCP surface. The CNT/Pt@CdSe-OCPs has a significantly superior $MA_{Pt}$ compared to that of Pt/C. In particular, at potential of -0.15 V *vs.* RHE (reference hydrogen electrode), the CNT/Pt@CdSe-OCPs display a $MA_{Pt}$ of 166.2 A mg$^{-1}$ and 42.0 A mg$^{-1}$ in 0.5 M $H_2SO_4$ and 1 M KOH, respectively. These $MA_{Pt}$ are 13.4 and 21.0 times those measured for Pt/C reference (12.4 in 0.5 M $H_2SO_4$ and 2.0 A mg$^{-1}$ in 1 M KOH). The turnover frequency ($TOF$) based on the Pt component for both CNT/Pt@CdSe-OCPs and Pt/C was calculated at various potentials (*i.e.*, 0.05, 0.10 and 0.15 V *vs.* RHE) from the current densities obtained from the LSV curves (see Supporting Information, Equation (S4)). As shown in **Figure 3b**, the $TOF$s based on Pt component of CNT/Pt@CdSe-OCPs are significantly higher than those of Pt/C at identical potentials. For example, at potential of –0.1 V *vs.* RHE, the $TOF$s of CNT/Pt@CdSe-OCPs are 84.2 s$^{-1}$ and 14.2 s$^{-1}$ in 0.5 M $H_2SO_4$ and 1 M KOH, respectively. These $TOF$s are respectively ~ 11 and 12 times those of Pt/C (7.5 s$^{-1}$ and 1.2 s$^{-1}$ in 0.5 M $H_2SO_4$ and 1 M KOH, respectively). To further understand the significant HER-activity of the Pt component in CNT/Pt@CdSe-OCPs, the HER kinetics were evaluated by Tafel plot analysis (see Supporting Information, Equation (S5))[97,98] (**Figure 3c**). The slope in Tafel plots reflects the rate-determining step (RDS) of the HER.[99,100] Briefly, the HER proceeds through the so-called Volmer step, leading to the formation of atomic hydrogen adsorbed on the electrocatalyst surface ($H_{ads}$) (*i.e.*: $H_3O^+ + e^- \rightleftarrows H_{ads} + H_2O$ in acidic media; $H_2O + e^- \rightleftarrows H_{ads} + OH^-$ in alkaline media), followed by either an electrochemical Heyrovsky step (*i.e.*: $H_{ads} + H_3O^+ + e^- \rightleftarrows H_2 + H_2O$ in acidic media; $H_2O + H_{ads} + e^- \rightleftarrows H_2 + OH^-$ in alkaline media) or a chemical Tafel recombination step (*i.e.*, $2H_{ads} \rightleftarrows H_2$).[74,97,99] For an insufficient $H_{ads}$ surface coverage, the Volmer reaction is the RDS of the HER, and a theoretical Tafel slope of 120 mV dec$^{-1}$ is observed.[99,100] Instead, for a borderline case in which there is a high $H_{ads}$ surface coverage (*i.e.*, $\Delta G_{Hads}^0$ is close to zero), the HER-kinetic is dominated by the Heyrovsky or Tafel reaction, and a Tafel slope of 40 mV dec$^{-1}$ or 30 mV dec$^{-1}$ is detected.[99,100] Intermediate Tafel slope values between 40 mV dec$^{-1}$ and 120 mV dec$^{-1}$, as well as potential-dependent Tafel slopes, can be experimentally observed due to the potential-dependence $H_{ads}$ surface coverage, which can span values between those assumed to model the HER reaction pathways.[101,100] The CNT/Pt@CdSe-OCPs shows a Tafel slope of 61.3 mV dec$^{-1}$ and 106.0 mV dec$^{-1}$ in 0.5 M $H_2SO_4$ and 1 M KOH, respectively. This indicates that the HER proceed through the Volmer-Heyrovsky mechanism. Instead, Pt/C shows a Tafel slope of 29.8 mV dec$^{-1}$ and 39.8 mV dec$^{-1}$, in 0.5 M $H_2SO_4$ and 1 M KOH. These values suggest that the Tafel recombination step is the RDS in acidic conditions, while the HER kinetic in 1 M KOH is slowed down due to limited efficiency of Pt to perform the Volmer step in alkaline media (*i.e.*, to dissociate the $H_2O$).[102] Overall, the significant differences of the HER kinetics observed among the electrodes suggest that the bonding of isolated nm-size Pt species onto CNT/Pt@CdSe-OCPs can alter the RDS typically observed in commercial Pt/C. This is somehow expected, as isolated active sites cannot, intrinsically, perform HER through a Volmer-Tafel mechanism, since no $2H_{ads}$ can be nearby to chemically react and form $H_2$. Although the origin of the high HER-activity of CNT/Pt@CdSe-OCPs cannot be fully explained at this stage, our material characterizations provide significant hints. Specifically, XPS analysis revealed that Pt is manifested in a partially positive charged state, $Pt^{+\delta}$, which can result in a tuning of Gibbs free energy of hydrogen adsorption $\Delta G_{Hads}$ on its surface from slightly negative values[3,103] to the optimal thermo-neutral value (*i.e.*, $\Delta G_{Hads}$ = 0 eV).[104] The establishment of an electrochemical interaction between Cd chalcogenide octapods and Pt nanocrystals was previously demonstrated by ultrafast optical measurements, which evidenced that Pt domains can capture hot electrons faster than energy relaxation and Auger recombination.[77] The chemical state of Pt in our system may facilitate the $H_{ads}$ desorption and completing the HER, accelerating the HER at the Pt active sites.[21,57,86,87] In alkaline media, CdSe-OCP surface may also promote the $H_2O$ dissociation in alkaline media, since the $OH^-$ can be captured by positively charged Cd (*i.e.*, $Cd^{2+}$ in CdSe-OPCs).[57]

Beyond the catalytic activity of the CNT/Pt@CdSe-OCPs, the electrochemical performance retention over time is a fundamental practical requisite. Actually, the physical dissolution of Pt, as well as its redeposition-induced agglomeration (*i.e.*, ripening), are often deleterious for long-term applications of efficient Pt-based catalysts (including Pt/C) both in acidic[105] and alkaline media.[106] **Figure 3d** reports the chronoamperometry measurements (*i.e., j–t* curves) for the CNT/Pt@CdSe-OCPs in 0.5 M $H_2SO_4$ and 1 M KOH, respectively, together with those measured for Pt/C reference. A constant overpotential, providing a starting cathodic current density of 50 mA $cm^{-2}$, was applied over 20 h. In 0.5 M $H_2SO_4$, both the CNT/Pt@CdSe-OCP and the Pt/C exhibited a durable catalytic activity. The increase of the current density in CNT/Pt@CdSe-OCP may be ascribed to a strengthening of the interaction between Pt and CdSe-OCP surface under cathodic polarization. Alternatively, the increment of the HER-activity during the measurements might be due to favorable morphological changes of the electrode during the HER process. In fact, in our electrode, the Pt@CdSe-OCPs adhere to CNTs without the need of any kind of electrocatalyst binders, such as sulfonated tetrafluoroethylene-based fluoropolymer-copolymers (*e.g.*, Nafion). Therefore, mechanical stresses originated by $H_2$ bubbling could cause a reorientation of the Pt@CdSe-OCPs, which progressively increase the $H^+$ accessibility to the HER-active sites of nanostructured electrodes.[107,108] Prospectively, the use of electrocatalyst binders could "freeze" an optimized electrode morphology, preferably obtained during the deposition of the electrode films and controllable by the presence of surface ligands.[64] Further studies are still needed to clarify the origin of this beneficial behaviour. In 1 M KOH, the CNT/Pt@CdSe-OCP almost retains its initial performance
(-2.8% after 20 h), while the Pt/C significantly decreases its current density over time (–63.8% after 20 h). This means that the hierarchical structure of the CNT/Pt@CdSe-OCPs, as well as the chemical interaction between Pt and CdSe surface (including ligands), are effective for the immobilization of nm-size Pt species, which do not dissolve or coalesce during the HER process. The catalytic properties of the CNT/Pt@CdSe-OCP were further examined for oxygen evolution reaction (OER) (**Figure S11**), showing a remarkable catalytic activity superior to those of both CdSe-OCPs and Pt/C. In our view, the origin of the OER-activity of the CNT/Pt@CdSe-OCPs is associated to the synergistic effects of multiple components, similarly to those often reported for heterogeneous catalysts.[57,74] However, chronoamperometry measurements evidences an insufficient durability of the OER-activity of the CNT/Pt@CdSe-OCPs over a time-of-hour scale (although superior to the one of Pt/C). Therefore, the CNT/Pt@CdSe-OCPs currently unfit practical requirements for OER.

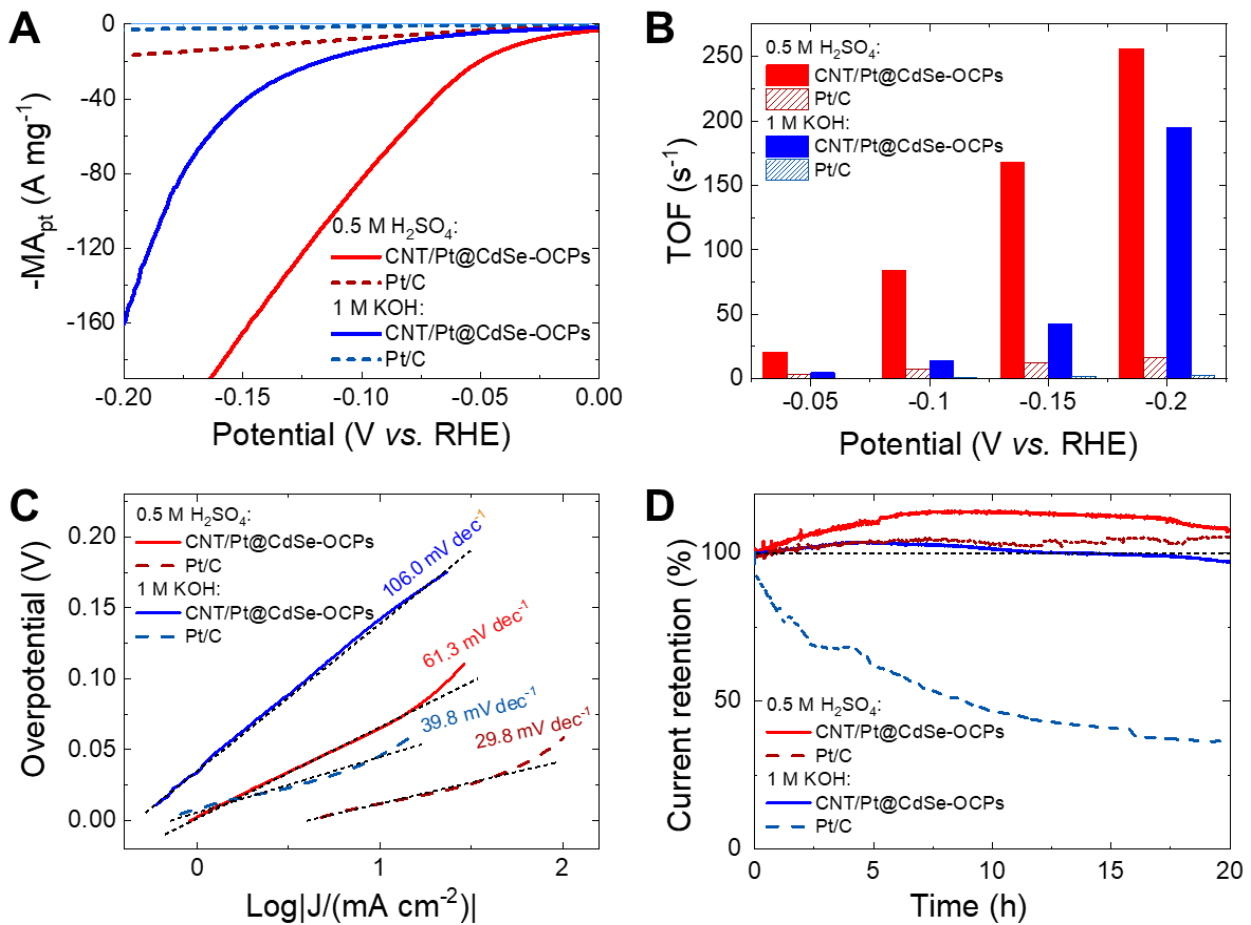

**Figure 3.** (A) Cathodic LSV scans (reported as $-MA_{Pt}$ vs. potential) for CNT/Pt@CdSe-OCPs and Pt/C in 0.5 M $H_2SO_4$ or 1 M KOH. (B) Comparison between the *TOF*s of the Pt component in 0.5 M $H_2SO_4$ or 1 M KOH at different overpotentials. (C) Tafel plots for CNT/Pt@CdSe-OCPs and Pt/C in 0.5 M $H_2SO_4$ or 1 M KOH. The linear fits of the curves are also shown, together with the extrapolated Tafel slopes. (D) Stability tests (chronoamperometry measurements) for the CNT/Pt@CdSe-OCPs and Pt/C operating in 0.5 M $H_2SO_4$ or 1 M KOH over 20 h at a constant overpotential, corresponding to a starting cathodic current density of 50 mA cm$^{-2}$.

## CONCLUSION

In summary, we have reported the ability of colloidally synthetized branched nanocrystals to host and immobilize isolated metal nanocrystals, aiming to tune the realization of advanced hierarchical multi-component catalysts for water splitting, especially for HER. We have directly bound Pt onto the sidewalls of the arms of CdSe octapods, resulting in a hierarchical dual-component system (Pt@CdSe-OCPs). The chemical interaction between CdSe-OCPs and Pt allows ultralow amount of Pt (in the order of few µg cm$^{-2}$) to be immobilized onto CdSe surface, preventing Pt coalescence during either synthesis or catalytic processes. By taking advantage of their branched structure, Pt@CdSe-OCPs self-assemble in a network when deposited on a CNT buckypaper used as the current collector. In addition, the as-formed network strongly adhere to the CNT buckypaper without resorting to any binding agent, since the pods effectively grab the CNT tangle. The hierarchical structure of the as-produced electrode (CNT/Pt@CdSe-OCPs) maximizes the utilization efficiency of the Pt, as a consequence of the optimal surface-to-bulk atomic ratio of the Pt. Meanwhile, the chemical interaction between Pt and the CdSe surface improve the catalytic properties of metallic Pt, without incurring in the common durability issues often reported for catalysts adopting ultralow Pt mass loadings (*e.g.*, Pt dissolution/deactivation effects). At potential of –0.15 V *vs.*

RHE, the proposed CNT/Pt@CdSe-OCP show Pt mass activity more than an order of magnitude higher compared to the Pt/C benchmark (13.4 and 21.0 times higher in 0.5 M $H_2SO_4$ and in 1 M KOH, respectively). The $MA_{Pt}$ of the CNT/Pt@CdSe-OCP corresponds to a turnover frequency based on Pt component as high as 168.0 s$^{-1}$ and
42.4 s$^{-1}$ in 0.5 M $H_2SO_4$ and 1 M KOH, respectively. The CNT/Pt@CdSe-OCP also exhibits a long-term durability for HER (> 20 h) in both the investigated media without using any polymeric binding agent, while the Pt/C reference progressively degrades in alkaline condition. These results experimentally prove that the branched colloidal nanocrystals can be exploited for the development and the theoretical study of a novel class of efficient and durable heterogeneous noble-metal-based catalysts.

## AUTHOR INFORMATION

**Corresponding Author**

* Tel.: +39 010 71781795. E-mail:Francesco.bonaccorso@iit.it

**Funding Sources**

This project has received funding from the European Union's Horizon 2020 research and innovation program under grant agreement No. 785219—GrapheneCore2.

## ACKNOWLEDGMENT

We thank Lea Pasquale and Mirko Prato (Materials Characterization Facility – Istituto Italiano di Tecnologia) for the support in the XRD and XPS data acquisition, respectively; the Electron Microscopy facility – Istituto Italiano di Tecnologia for the support in the TEM/EDS data acquisition; the Clean Room facility – Istituto Italiano di Tecnologia for the access to carry out the SEM/EDS characterizations and the Smart Materials facility – Istituto Italiano di Tecnologia for the access to carry out FTIR characterization.

## Supporting Information

**Supplementary Experimental Section**

**Chemicals**. Copper(II) acetylacetonate (Cu(acac)$_2$, 99.99%), cadmium chloride (CdCl$_2$, 99.99%), 1-hexanethiol (HT, 95%), 1-dodecanethiol (DDT, ≥ 98%), 1-octadecanethiol (ODT, 98%), oleylamine (OLAM, 70%), octadecylphosphonic acid (ODPA, 97%), 1,2-hexadecanediol (90%), oleic acid (OA, 90%), diphenyl ether (99%), 1,2-dichlorobenzene (99%) and Pt(II) acetylacetonate (Pt(acac)$_2$, ≥ 99.98%) were purchased from Sigma-Aldrich. Selenium powder (Se, 99.99%), cadmium oxide powder (CdO, 99.999%), tri-*n*-octylphosphine oxide (TOPO, 99%), and tri-*n*-octylphosphine (TOP, 97%) were purchased from Strem Chemicals. Hexylphosphonic acid (HPA) was purchased from Polycarbon Industries. All the above chemicals were used as received and all the syntheses were carried out using a standard Schlenk line. Sulfuric acid (H$_2$SO$_4$) (99.999%), potassium hydroxide (KOH) (≥ 85% purity, ACS reagent, pellets), platinum on carbon (Pt/C) (10 wt.% loading) and Nafion solution (5 wt.%) were purchased by Sigma Aldrich. Carbon nanotubes (single-/double-walled carbon nanotubes, > 95% purity) were purchased from Cheap Tubes.

**Synthesis of Cu$_{2-x}$Se seeds.**[1,2] In a three-neck flask 1 mmol of Cu(acac)$_2$ (262 mg) was dissolved in 9.5 mL of OLAM and 3 ml of DDT and degassed for 1h at 60 °C under stirring. Then, the temperature was increased to 220 °C under N$_2$ flow. Once the solution became orange-brown, 1 mL of a previously prepared solution of Se (1 M in OLAM-DDT mixture, 50-50 % vol.) and 1.5 mL of degassed DDT were injected in the flask. After 4 min, the flask was cooled to room temperature and the solution was transferred to a N$_2$ filled glovebox to proceed with multiple washing cycles *via* precipitation with methanol and re-dissolution in toluene to eliminate the excess of ligands from the synthesis. The Cu$_{2-x}$Se seeds were then stored in the N$_2$-filled glove box as sealed toluene dispersions and used in the synthesis of octapods.

**Syntheses of CdSe-OCPs and CdSe quantum dots (QDs).**[3,4] In a three-neck flask 60 mg of CdO, 6 mg of CdCl$_2$, 3 g of TOPO, 290 mg of ODPA, and 80 mg of HPA were degassed for 1h at 130 °C under stirring. The solution was heated up to 350 °C under N$_2$ flow. When the solution became transparent (~ 260 °C), 2.5 mL of anhydrous TOP were injected in the flask. In the synthesis of CdSe-OCPs, once the temperature reached 350 °C, 200 μL of the Cu$_{2-x}$Se seeds previously prepared (concentration ~6.4 μM), 500 μL of TOP, and 620 μL of Se precursor (32 mg mL$^{-1}$ solution in TOP) were injected in the mixture. When synthesizing CdSe-QDs, no Se precursor was added. The reaction was run for 10 min before cooling it down to room temperature. When the temperature dropped below 100 °C, 3 mL of anhydrous toluene were added to the resulting product. The dispersion was then washed twice using a mixture of ethanol and methanol and centrifuged to remove the excess of free-ligands that can induce particle aggregation.

**Debundling of CNTs.** The CNTs were dispersed in N-methyl-2-pyrrolidone (NMP) with a concentration of 0.2 g L$^{-1}$ by means of ultrasonication-based de-bundling.[5,6] In particular, 10 mg of CNT powder was added to 50 mL of NMP. The dispersion was then sonicated for 30 min by using a sonic tip (Vibra-cell 75185, Sonics) with vibration amplitude set to 45%. The sonic tip was pulsed for 5 s ON and 2 s OFF to reduce the solvent heating, which was also mitigated by an ice bath around the beaker.

**Characterization of the materials.** Fourier-transform infrared spectroscopy (FTIR) measurements were carried out using a Bruker Vertex® 70v in attenuated total reflection (ATR) mode, in order to evaluate the surface of the Pt@CdSe-OCPs before and after the purification step performed prior the electrochemical characterization. The ATR-FTIR spectra were acquired over 4000–550 cm$^{-1}$ operating range, with a resolution of 4 cm$^{-1}$ and averaging 100 scans for each spectrum. The samples were prepared by drop-casting an aliquot of the as-prepared or purified Pt@CdSe-OCP dispersions on the surface of the ATR crystals. The purified Pt@CdSe-OCP dispersions were obtained by washing the as-prepared dispersions with isopropyl alcohol (IPA) with a Pt@CdSe-OCP dispersion:IPA volume ratio of 1:100, followed by centrifugation and material redispersion in toluene. The washing treatment simulates the one used for the investigated electrodes to remove the residual ligands of Pt@CdSe-OCPs, as well as residuals of NMP, which was used to formulate the CNT dispersion. The FTIR measurements were performed on dried samples.

Bright-field transmission electron microscopy (BFTEM) images were acquired with a JEOL JEM 1011 TEM (thermionic W filament), operating at 100 kV. Morphological and statistical analyses were performed using ImageJ software (NIH) and OriginPro 9.1 software (OriginLab), respectively. Samples for the BFTEM measurements were prepared by drop casting the as-prepared material dispersions onto ultrathin carbon-on-holey carbon film/Cu grid (UHC/Cu) and rinsed with deionized water and subsequently dried overnight under vacuum.

High resolution TEM (HRTEM), high-angle annular dark-field imaging scanning TEM (HAADF-STEM) imaging and energy dispersive X-ray spectroscopy in scanning TEM (STEM-EDS) analyses were carried out on a JEOL JEM-2200FS TEM (Schottky emitter), operating at 200 kV, equipped with a CEOS corrector for the objective lens, an in-column image filter (Ω-type) and Bruker XFlash 5060 EDS silicon-drift detector. Samples for HRTEM, HAADF-STEM and STEM-EDS were prepared by casting the as-prepared Pt@CdSe-OCP dispersion onto an ultrathin carbon-on-holey carbon film/Cu grid (UHC/Cu). The samples were dried under vacuum overnight before the measurements. The presented STEM-EDS elemental maps were obtained by integration of the $K_\alpha$ peaks for Se and $L_\alpha$ peaks for Cd and Pt.

X-ray photoelectron spectroscopy (XPS) analysis was accomplished on a Kratos Axis UltraDLD spectrometer at a vacuum better than $10^{-8}$ mbar, using a monochromatic Al Kα source operating at 20 mA and 15 kV and collecting photoelectrons from a 300 × 700 µm$^2$ sample area. The charge compensation device was not used. Wide spectra were acquired at pass energy of 160 eV and energy step of 1 eV, while high-resolution spectra of Cd 3$d$, Se 3$d$, Pt 4$f$, O 1$s$ and C 1$s$ peaks were acquired at pass energy of 10 eV and energy step of 0.1 eV. The samples were prepared by drop-casting the CdSe-OCP or Pt@CdSe-OCP dispersions on an Au-coated Si chip. Calibration of the binding energy was performed by setting the C 1s peak to 284.8 eV. Gold 4$f$ peaks were not observed due to high substrate coverage. Data analysis is carried out with CasaXPS software (version 2.3.19PR1.0). Shirley backgrounds and Gaussian/Lorentzian (GL) product line shapes with 30% Lorentzian weight were generally used. Cadmium 3$d$ peaks were fitted with 70% Lorentzian weight. The fit of Pt 4$f$ peaks with asymmetric functions used the LA(1, 1.3, 200) line shape of CasaXPS.[7] Subtraction of the CdSe-OCP signal from the Pt 4$f$ signal of Pt@CdSe-OCP was carried out with the following method: the CdSe-OCP signal between 82.5 eV and 64.5 eV was smoothed by polynomial regression (degree 35), scaled to match the Pt 4$f$ pre-edge (B.E. = 66.5 eV) intensity measured for Pt@CdSe-OCPs and then subtracted from the Pt@CdSe-OCP spectrum.

X-ray diffraction (XRD) patterns were acquired with a PANalytical Empyrean X-ray diffractometer equipped with a 1.8 kW CuKα ceramic X-ray tube, a PIXcel3D 2×2 area detector and operating at 45 kV and 40 mA. The samples for XRD were prepared by drop-casting synthetized material dispersions onto Si/SiO2 substrates and dried overnight under vacuum.

Elemental analysis of the Pt@CdSe-OCPs was performed by inductively coupled plasma optical emission spectroscopy (ICP-OES) and inductively coupled plasma optical emission mass spectroscopy (ICP-MS). The ICP-OES measurements were carried with a ThermoFisher ICAP 6000 Duo inductively coupled plasma optical emission spectrometer. The analyzed material dispersions were prepared in a 25 mL volumetric flask. Each sample was obtained by digesting 25 µL of the as-produced Pt@CdSe-OCP dispersion in 2.5 mL of aqua regia overnight and then diluting the solution to 25 mL with Millipore water. Prior to analysis, the samples were stirred by vortex at 2400 rpm for 10 s and filtered using a PTFE membrane (0.45 µm pore size). The ICP-MS measurements were carried out using a iCAP-TQs ThermoFisher ICP-MS. $^{209}$Bi and $^{73}$Ge were used as Internal Standard for the $^{195}$Pt the $^{112}$Cd, respectively. The plasma power was set to 1450W. The instrument was used in Triple Quadrupole Mode and the He was used as collision gas in the quadrupole cell. The sample was prepared by diluting 1,000,000 times the as-produced Pt@CdSe-OCP dispersion with 10 mL H2O/HNO₃ (9:1).

**Characterization of the electrodes.** Scanning electron microscopy (SEM)/energy dispersive X-ray spectroscopy (EDS) characterization of the as-produced electrodes was performed using a Helios Nanolab® 600 DualBeam microscope (FEI Company) equipped with an X-Max detector and INCA® system (Oxford Instruments) operating at 5 kV and 0.2 nA as measurement conditions for imaging and 15 kV and 0.8 nA for EDS analysis. The samples were imaged without any metal coating or pre-treatment.

Electrochemical characterization was performed at room temperature in a flat-bottom quartz cell using the three-electrode configuration of a CompactStat potentiostat/galvanostat station (Ivium), controlled *via* Ivium's own IviumSoft software. A glassy carbon rod and a KCl-saturated Ag/AgCl electrode were used as counter-electrode and reference electrode, respectively. The measurements were carried out in 200 mL of 0.5 M $H_2SO_4$ or 1 M KOH MilliQ® solutions for acidic and basic conditions, respectively. Before starting the measurements, dissolved $O_2$ was purged from the electrolyte by flowing $N_2$ gas throughout the liquid volume using a porous frit. The Nernst equation, in the following form, was used to convert the potential difference between the working electrode and the KCl-saturated Ag/AgCl reference electrode to the reversible hydrogen electrode (RHE) scale:

$E_{RHE} = E_{Ag/AgCl} + 0.059 \times pH + E^0_{Ag/AgCl}$ (S1)

where $E_{RHE}$ is the converted potential *vs.* RHE, $E_{Ag/AgCl}$ is the experimental potential measured against the saturated-KCl Ag/AgCl reference electrode, and $E^0_{Ag/AgCl}$ is the standard potential of Ag/AgCl at 25 °C (0.1976 V *vs.* RHE).

Linear sweep voltammetry (LSV) curves were acquired at 5 mV s$^{-1}$ potential scan rate and were *iR*-corrected (100% *iR*-drop compensation) by considering *i* as the measured working electrode current and *R* as the series resistance arising from the working electrode substrate and the electrolyte resistances.

Cyclic voltammetry (CV) measurements of the CNT buckypaper were carried out in a non-faradaic region at different potential scan rate (ranging from 10 to 400 mV s$^{-1}$). By plotting the difference between anodic and cathodic current densities ($\Delta j = (j_a - j_c)$) at 0.175 V *vs.* RHE as a function of the scan rate (*SR*), $C_{dl}$ can be calculated by:

$$C_{dl} = (1/2) \times (d(\Delta j)/d(SR))$$ (S2)

Electrochemical impedance spectroscopy (EIS), at open circuit potential and at 10 kHz, was used to measure *R*. The mass activity ($MA_{pt}$) and the turnover frequency (*TOF*) of the Pt was calculated according to the following equations:

$$MA_{Pt} = j/m_{Pt}$$ (S3)

$$TOF = i/(2F \cdot n)$$ (S4)

in which *j* is the measured current density, $m_{Pt}$ is the Pt mass loading (measured by ICP-MS analysis), *F* is the Faraday constant (96485.3 C mol$^{-1}$), and *n* is the number of Pt moles in the catalyst.

The linear portions of the Tafel plots (*i.e.*, overpotential *vs.* log(|*j*|) plot), as derived by *iR*-corrected LSV curves, were analysed by the fitting Tafel equation: [8,9]

$$\eta = b * |log(j)| + A$$ (S5)

where η is the overpotential *vs.* RHE, *b* is the Tafel slope and *A* is the intercept of the linear regression.

Stability tests were performed by chronoamperometry measurements, *i.e.,* by measuring the current in potentiostatic mode over time, at a fixed overpotential corresponding to a cathodic current density of 80 mA cm$^{-2}$.

## Scheme S1. Schematic illustration of the synthesis process for Pt@CdSe-OCPs

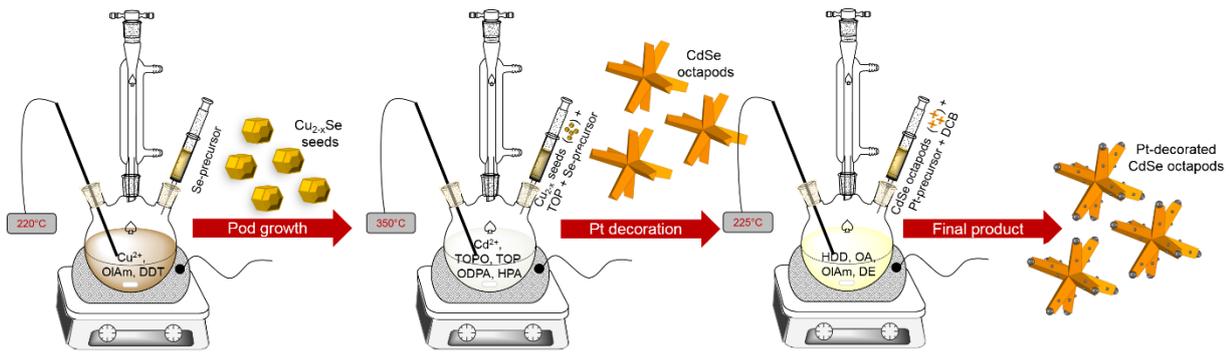

**Fourier-transform infrared spectroscopy analysis of Pt@CdSe-OCPs**

**Figure S1** shows the ATR-FTIR spectra of the as-synthetized (in red) and the purified Pt@CdSe-OCPs (in blue). The intensities of the FTIR bands attributed to the stretching modes of the P-O and P=O moieties in the region 1250 – 950 cm$^{-1}$ and CH$_2$ and CH$_3$ stretching modes from the alkyl chains in the region 3000 – 2800 cm$^{-1}$ can be used as a metric to evaluate the presence of the native (phosphorous-based) ligands involved in the synthesis protocol (*i.e.*, TOPO, HPA, and ODPA).[4] The FTIR spectra clearly show a significant reduction on the signal of the ligands after the purification step (see details in Supplementary Experimental Section).

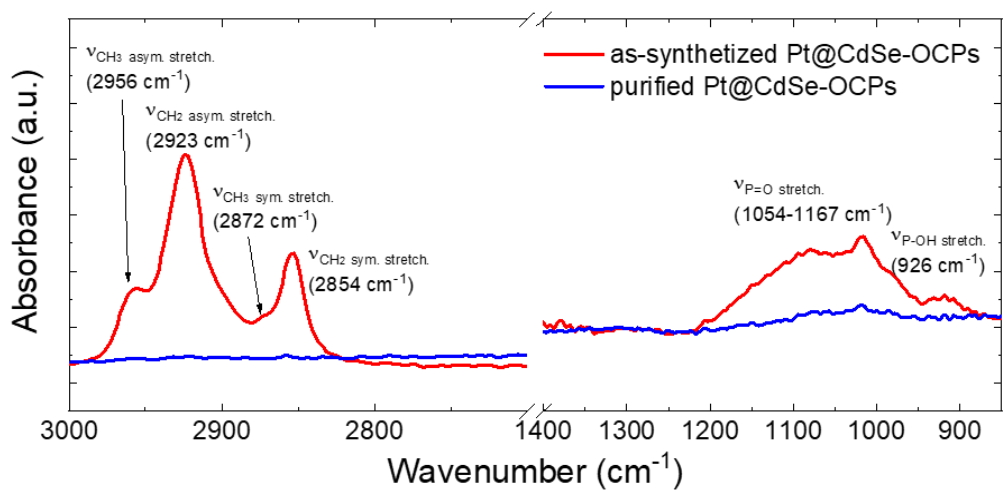

*Figure S1.* ATR-FTIR spectra of the as-synthetized Pt@CdSe-OCPs and the purified Pt@CdSe-OCPs.

## Transmission electron microscopy analysis of CdSe-OCPs

**Figure S2** shows representative BFTEM images of CdSe-OCP ensembles, whose statistical analysis allowed the relevant CdSe-OCP dimensions (the tip-to-tip arm length –L–, the arm diameter –D– and the actual arm length –$L_P$–) to be estimated.

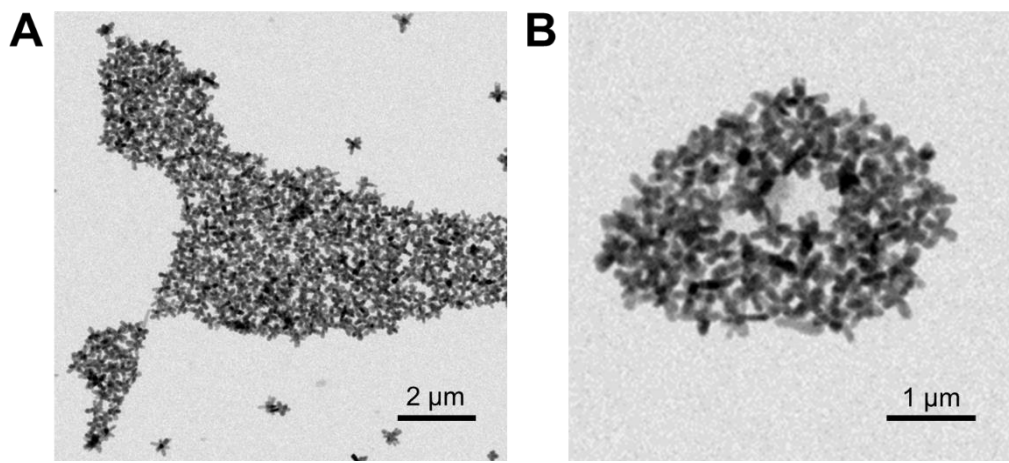

***Figure S2.*** *(A,B) BFTEM images of CdSe-OCP ensembles.*

**Figure S3** reports a HRTEM image of an isolated Pt@CdSe-OCP with the corresponding fast Fourier transform (FFT) acquired from different regions. The data indicate that the pods of Pt@CdSe-OCPs are wurtzite-type (hexagonal) CdSe (ICSD 415785), with <001> as elongation direction.

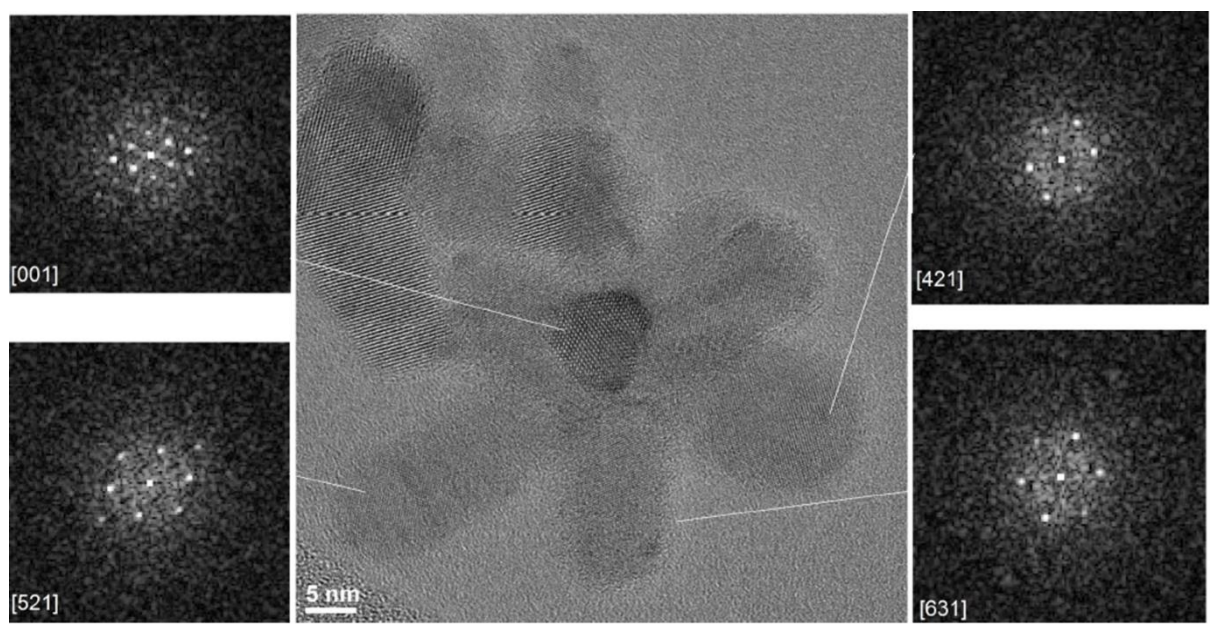

***Figure S3.*** *HRTEM image of an isolate Pt@CdSe-OCP with the corresponding fast Fourier transform (FFT) from different regions.*

## X-ray photoelectron spectroscopy analysis of CdSe-OCPs

**Figure S4** reports the XPS analysis of CdSe-OCPs. In the Cd 3$d$ spectrum (**Figure S4a**), the two peaks positioned at 405.4 and 412.2 eV are assigned to Cd 3$d_{5/2}$ and Cd 3$d_{3/2}$ bands, respectively, of the Cd$^{2+}$ state of the CdSe.[10] No peaks related to other oxidized states of Cd are observed. In the Se 3$d$ spectrum (**Figure 4b**), the bands peaking at 54.2 and 55.1 eV are attributed to the Se 3$d_{5/2}$ and Se 3$d_{3/2}$, respectively, of the Se moiety in CdSe. The peaks ascribed to a Se$^{4+}$ state in SeO$_2$ (typically located at 58.8 e),[11,12] are not observed. **Figure S4c** shows the XPS signal of CdSe-OCPs in the region of Pt 4$f$. This signal is assigned to the broad energy loss structure of CdSe-OCPs. The polynomial fit of this signal was used to model CdSe-OCP background in the Pt 4$f$ XPS spectrum of Pt@CdSe-OCPs.

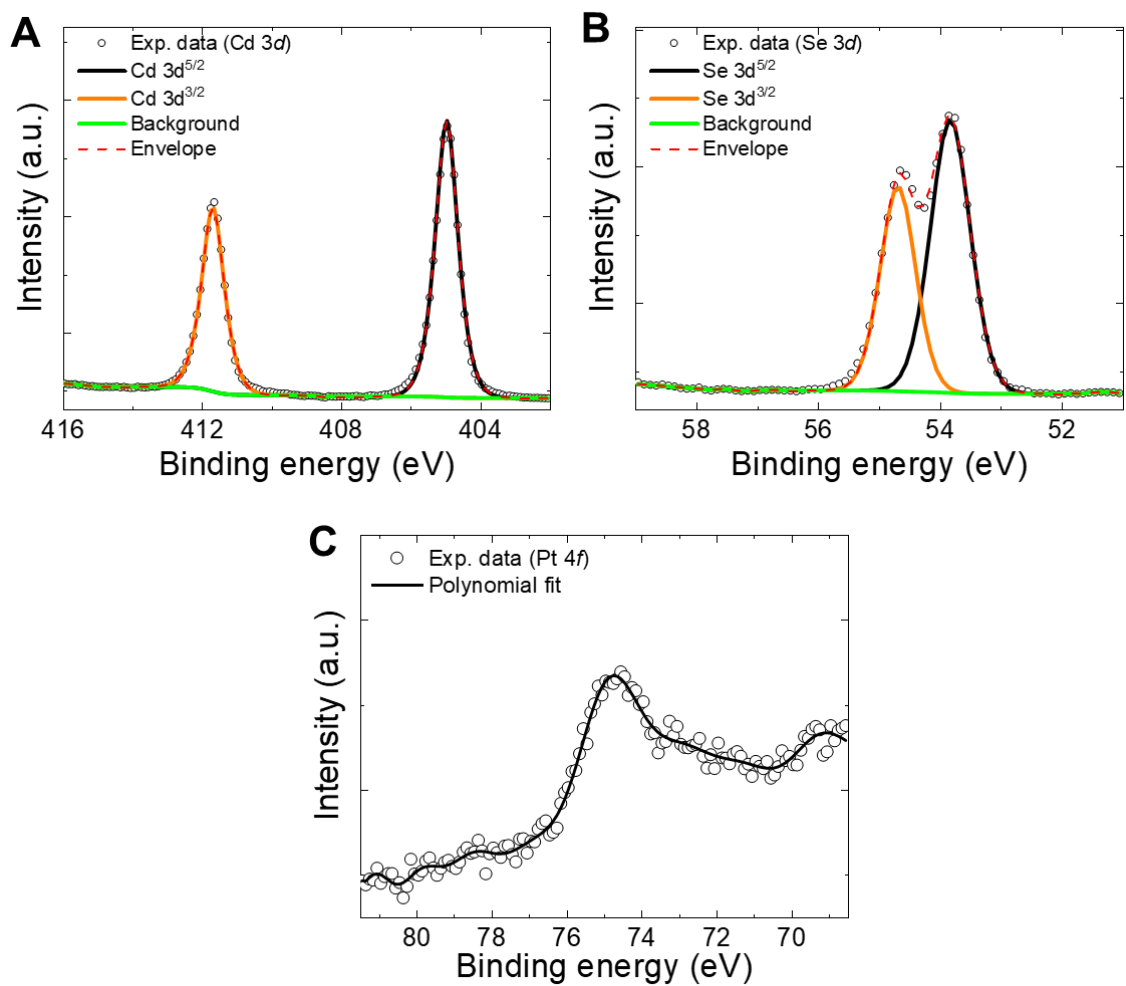

*Figure S4. (A) Cd 3d and (B) Se 3d XPS spectra of CdSe-OCPs, together with their deconvolutions. (C) XPS spectrum of CdSe in the region of Pt 4f. The polynomial fit of this signal was used to model the CdSe-OCP background in the Pt 4f XPS spectrum of Pt@CdSe-OCPs.*

## Supplementary X-ray photoelectron spectroscopy analysis of Pt@CdSe-OCPs

**Figure S5** reports the as-measured Pt 4*f* XPS spectrum of the Pt@CdSe-OCPs. The signal attributed to Pt element overlaps with the broad energy loss structure observed in CdSe-OCPs, as modeled by a polynomial fit (see **Figure S4c**). As shown in **Figure 1j** of the main text, the background signal ascribed to CdSe-OCPs was subtracted to the as-measured Pt 4*f* XPS spectrum in order to investigate the XPS signal assigned to the solely Pt chemical states.

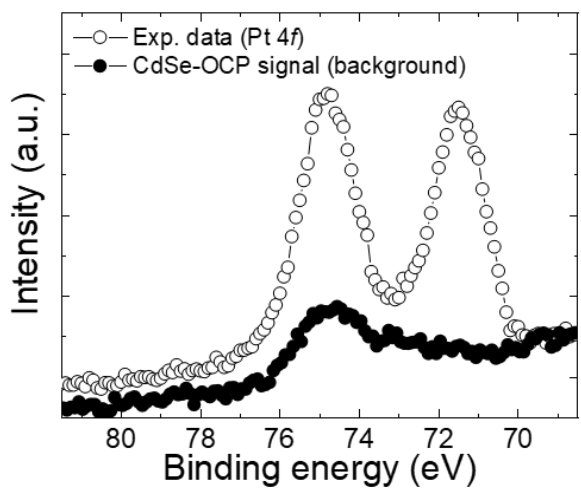

**Figure S5.** As-measured Pt 4f XPS spectrum of the Pt@CdSe-OCPs. The background attributed to the CdSe-OCPs, as observed in **Figure S4c**, is also plotted.

**Figure S6** shows Pt 4*f* XPS spectrum of the Pt@CdSe-OCPs deconvoluted with an asymmetric doublet. The resulting fit was considered alternatively to the optimal fit obtained using symmetric peaks and introducing a fraction (*ca.* 11%) of $Pt^{+\delta}$ contributing with a doublet peaking at higher binding energy (73.5 eV and 76.8 eV) compared to those of $Pt^0$ doublet.

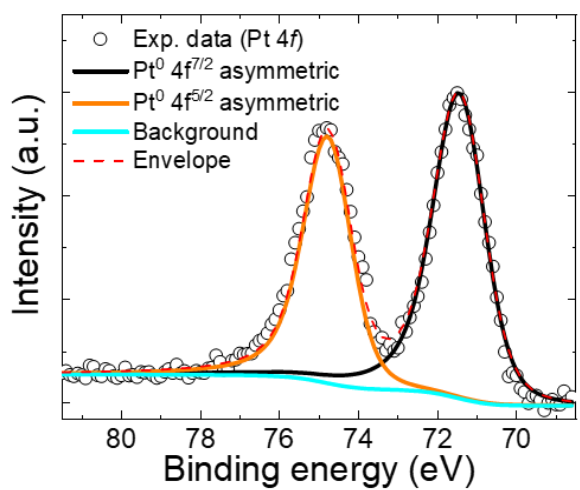

**Figure S6.** Pt 4f XPS spectra of Pt@CdSe-OCPs, together with its deconvolution using a single doublet with asymmetric peak shapes.

**Scanning electron microscopy analysis of the carbon nanotubes buckypaper**

**Figure S7** reports a top-view SEM image of a representative carbon nanotubes (CNT) buckypaper used as the catalyst support. The surface of the CNT buckypaper consists of a mesoporous network forming a bundle-like morphology.

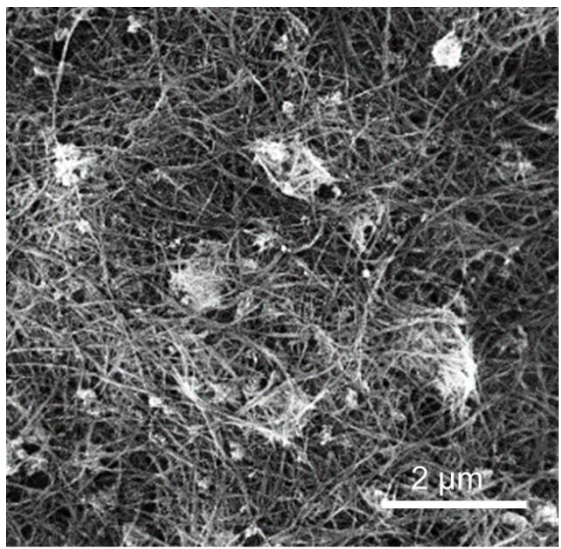

**Figure S7.** Top-view SEM image of a representative CNT buckypaper.

**Energy dispersive X-ray spectroscopy coupled with scanning electron microscopy analysis of CNT/Pt@CdSe-OCPs**

**Figure S8** shows the SEM-EDS analysis of CNT/Pt@CdSe-OCPs. The data indicate a Cd:Se atomic ratio of ~1.1. The excess of Cd agrees with XPS and ICP-OES/MS analyses reported in the main text. In addition, no Pt signal has been detect, confirming the ultralow Pt mass loading deduced by ICP-MS measurements and the Pt@CdSe-OCP mass loading (38 µg cm$^{-2}$).

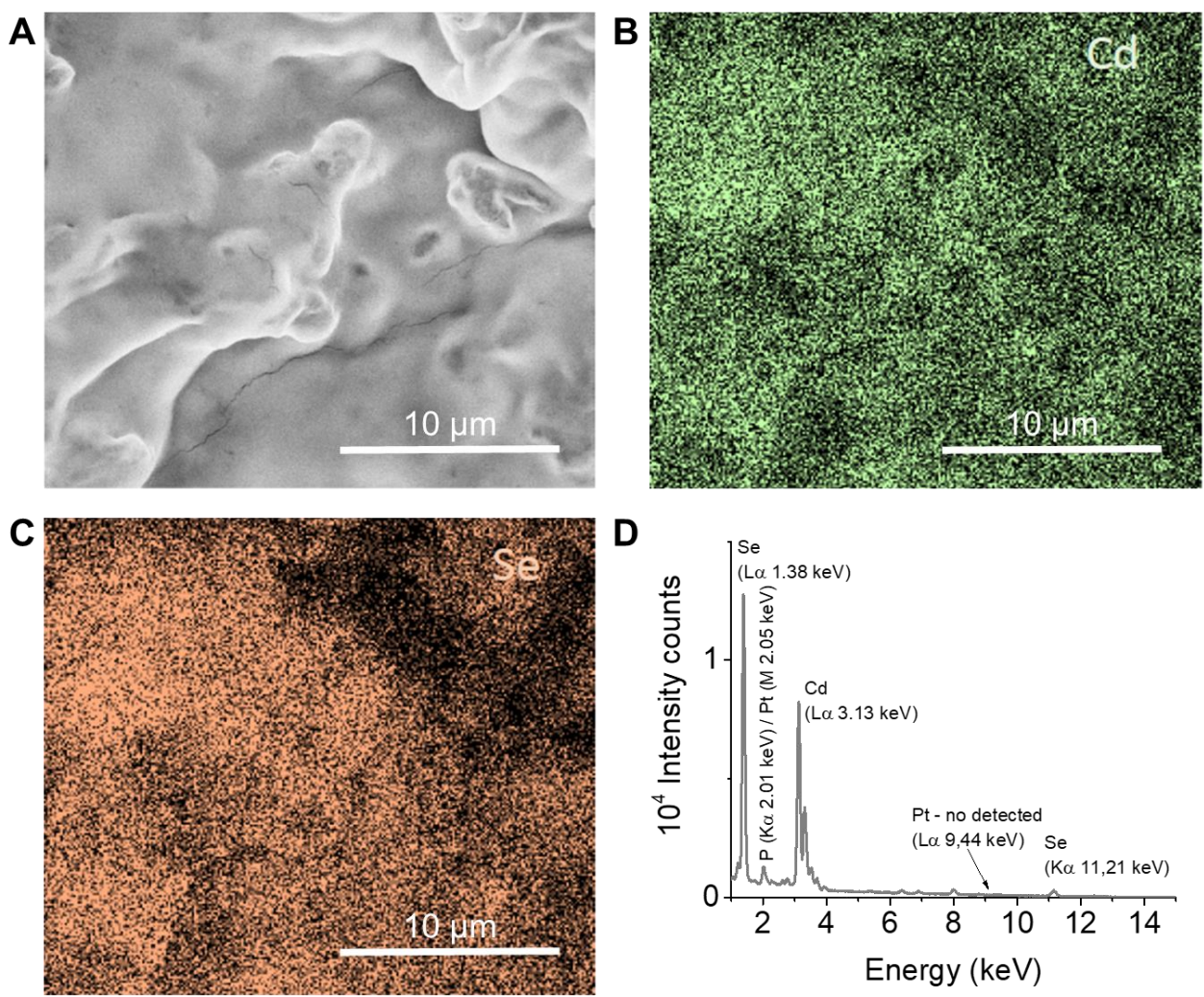

**Figure S8.** (A) Top-view SEM image of a CNT/Pt@CdSe-OCPs with the corresponding EDS maps for (B) Cd (Lα = 3.13 keV, green) and (C) Se (Lα = 1.38 keV, orange) and (D) the EDS spectrum.

## Cyclic voltammetry measurements of CNT buckypaper

**Figure S9a** shows the CV curves of CNT buckypaper measured at various potential *SR* (ranging from 10 to 80 mV s$^{-1}$) in a non-faradaic region of applied potentials (between 0.15 and 0.35 V *vs.* RHE). By plotting *Δj* at 0.175 V *vs.* RHE as a function of the potential *SR* (**Figure S9b**), the $C_{dl}$ can be estimated by using Equation (S2). By considering the potential *SR*s < 80 mV s$^{-1}$, the resistive losses can be neglected and the estimated $C_{dl}$ is ~790 mF cm$^{-2}$. This value indicates that the $C_{dl}$ of CNT buckypapers impedes a reliable evaluation of the $C_{dl}$ of our catalytic films deposited onto CNT buckypapers.

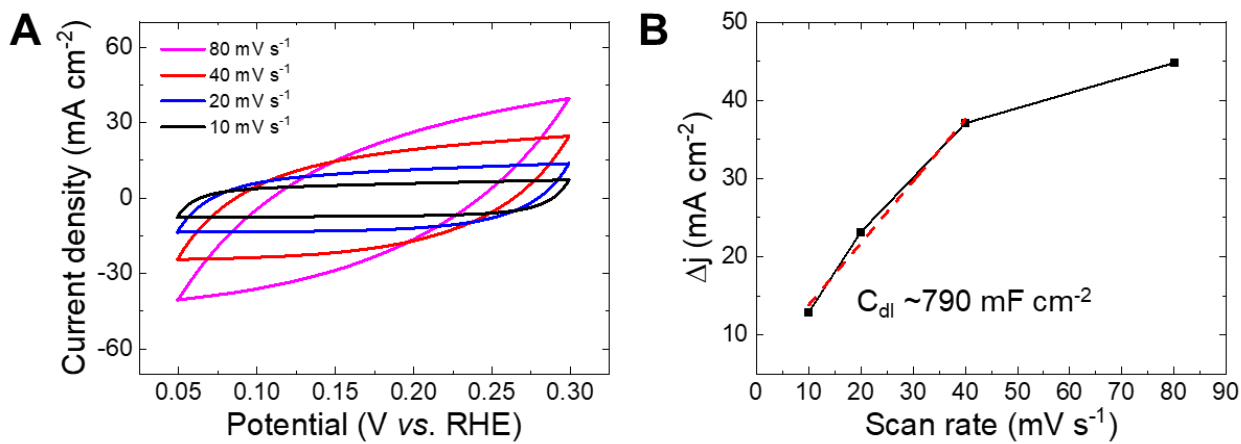

**Figure S9.** (A) CV measurements at various potential SRs and (B) potential SR dependence of the Δj (at 0.175 V vs. RHE) for CNT buckypaper.

## Linear sweep voltammetry analysis of CNT/CdSe-OCPs

**Figure S10** shows the cathodic LSV scan measured for CNT/CdSe-OCPs in both 0.5 M $H_2SO_4$ and 1 M KOH. The data indicate that the CNT/CdSe-OCPs does not exhibit relevant HER-activity in the potential range examined for the CNT/Pt@CdSe-OCPs (between –0.2 and 0 V *vs.* RHE, see **Figure 3** in the main text).

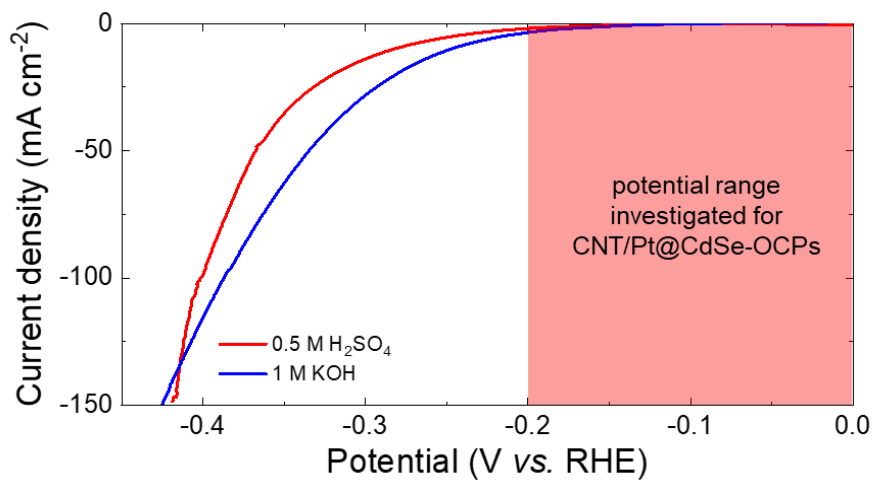

**Figure S10.** Cathodic LSV scans for the CNT/CdSe-OCPs in 0.5 M $H_2SO_4$ and 1 M KOH. The potential range examined for the CNT/Pt@CdSe-OCPs is also indicated.

## Linear sweep voltammetry analysis of the CNT/Pt@CdSe-OCP for oxygen evolution reaction

**Figure S11** reports the anodic LSV scans for CNT/CdSe-OCPs, CNT/Pt@CdSe-OCPs and Pt/C in 0.5 M $H_2SO_4$ (**Figure S11a**) and 0.5 M $H_2SO_4$ (**Figure S11b**). The data show that the CNT/Pt@CdSe-OCPs displays the highest oxygen evolution reaction (OER)-activity. This means that the Pt decoration of the CdSe-OCPs is effective to improve the catalytic performance of the CdSe-OCPs, which also exhibit a significant OER-activity. However, chronoamperometry tests, performed at potential corresponding to an initial anodic current density of 25 mA $cm^{-2}$, indicate an insufficient durability of the OER-activity of the CNT/Pt@CdSe-OCP over a time-of-hour scale (although superior to that of Pt/C) (**Figure S11c**).

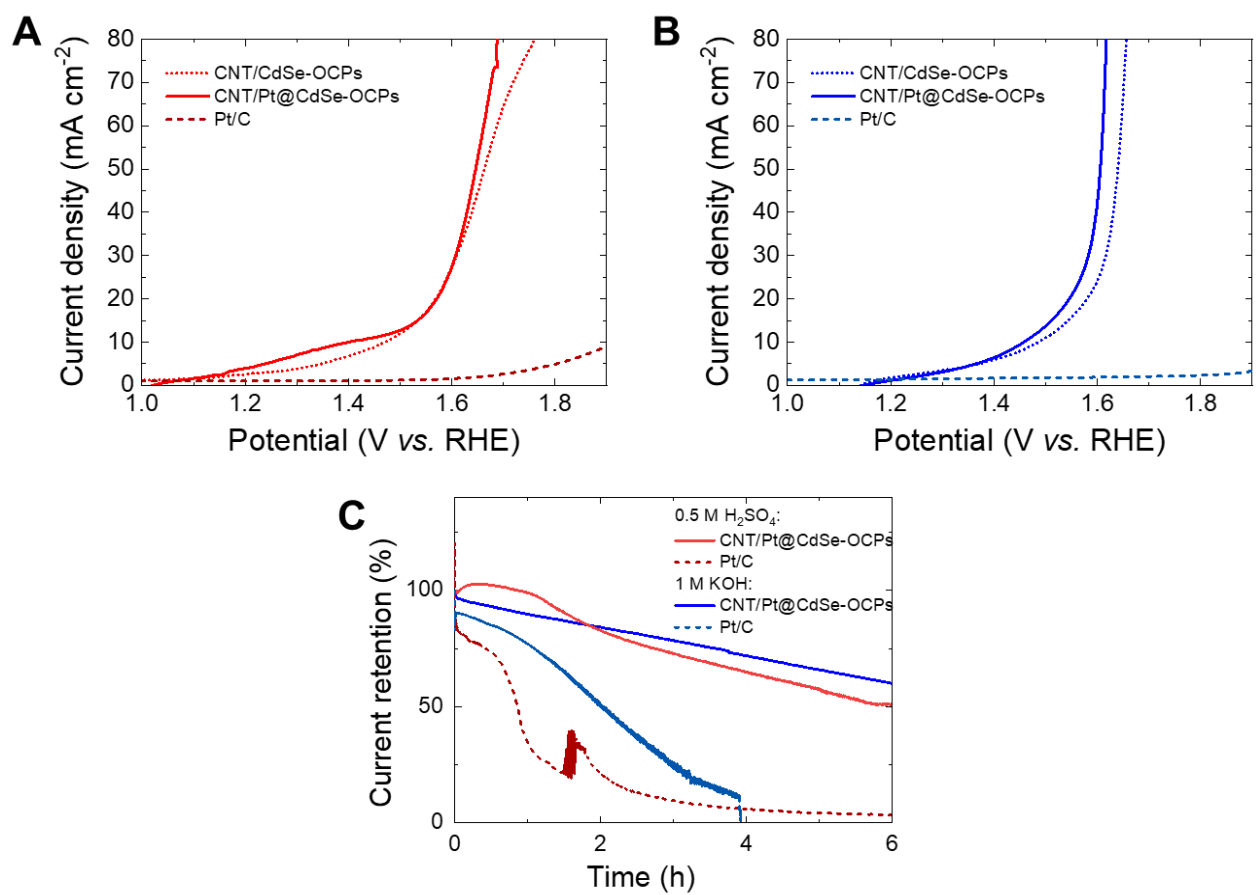

**Figure S11.** Anodic LSV scans for the CNT/CdSe-OCPs, CNT/Pt@CdSe-OCPs and Pt/C in (A) 0.5 M $H_2SO_4$ and (B) 1 M KOH. (C) Stability tests (chronoamperometry measurements) for the CNT/Pt@CdSe-OCPs and Pt/C operating in 0.5 M $H_2SO_4$ or 1 M KOH over 6 h at a constant overpotential, corresponding to a starting anodic current density of 25 mA $cm^{-2}$.